# Fluctuating nanomechanical systems in a high finesse optical microcavity.

*Ivan Favero[1), 2)]\*, Sebastian Stapfner[1)], David Hunger[1)], Philipp Paulitschke[1)], Jakob Reichel[3)], Heribert Lorenz[1)], Eva M. Weig[1)], Khaled Karrai[1)]*

\*ivan.favero@univ-paris-diderot.fr

1) CeNS and Fakultät für Physik, Ludwig-Maximilians-Universität, Geschwister Scholl-Platz 1, 80539 München, Germany.

2) Laboratoire Matériaux et Phénomènes Quantiques, Université Paris-Diderot, CNRS, UMR 7162, 10 rue Alice Domon et Léonie Duquet, 75013 Paris, France.

3) Laboratoire Kastler Brossel, Ecole Normale Supérieure, Université Pierre et Marie Curie, CNRS, 24 rue Lhomond, 75005 Paris, France.

**Confining a laser field between two high reflectivity mirrors of a high-finesse cavity can increase the probability of a given cavity photon to be scattered by an atom traversing the confined photon mode[1]. This enhanced coupling between light and atoms is successfully employed in cavity quantum electrodynamics experiments and led to a very prolific research in quantum optics[2,3]. The idea of extending such experiments to sub-wavelength sized nanomechanical systems has been recently proposed[4] in the context of optical cavity cooling[5,6]. Here we present an experiment involving a single nanorod consisting of about $10^9$ atoms precisely positioned to plunge into the confined mode of a miniature high finesse Fabry-Pérot cavity. We show that the optical transmission of the cavity is affected not only by the static position of the nanorod but also by its vibrational fluctuation. While an imprint of the vibration dynamics is directly detected in the optical transmission, back-action of the light field is also anticipated to quench the nanorod Brownian motion[4]. This experiment shows the first step towards optical cavity controlled dynamics of mechanical nanostructures and opens up new perspectives for sensing and manipulation of optomechanical nanosystems.**

Optomechanical systems combine an optical cavity and a (typically macro- or micron-scale) mechanical oscillator in a single device, leading to an increased coupling between them[7-14]. They have recently advanced into the fields of precision displacement measurement[15], investigation of mechanical systems close to their quantum-ground state[16], non-linear dynamics[17], or sensing applications[18]. In these systems, mechanical oscillators of various sizes are now being used: from a centimetre scale movable mirror[12] down to a 100 nm-sized beam integrated in a microwave strip line[19]. At the same time, nanomechanical sytems are being optically controlled without the use of a cavity[20,21]. Here we investigate a system for cavity nano-optomechanics in the optical domain, where a 100 nm diameter vibrating nanorod is coupled to a high finesse optical micro-cavity of small mode volume. For our experiment, we choose carbon-based nanorods which combine a large mechanical stiffness, low mass and flexible device preparation. Each nanorod is grown by Electron-Beam Deposition (EBD)[22] at the front extremity of a silicon microlever used for atomic force microscopy (AFM). As displayed in Fig. 1a the nanorod extends along the lever axis, as opposed to conventional AFM cantilever probes with the tip perpendicular to the lever plane. The shape of the nanorods is slightly conical (Fig. 1b) with lengths ranging from 3 to 5 μm and mean diameters between 80 and 160 nm. First, we investigate the mechanical properties of the rods by ultrasonic actuation. To this end, the silicon lever hosting the nanorod is mounted on a high frequency piezo-transducer and placed in the chamber of a Scanning Electron Microscope (SEM) at a pressure of about $10^{-6}$ mbar. With a frequency generator (RS SML02), the piezo-transducer is excited to actuate the oscillator. The mechanical response of the microlever-nanorod system is simultaneously investigated by monitoring the envelope of the resulting oscillation using the SEM video images[23]. Around 13 kHz, the first flexural mode of the host silicon lever is found in compliance with its specs, with a vibration amplitude of a few tens of microns at resonance. Other modes of the microlever are observed up to 300 kHz. At higher frequency, the lowest vibrational resonance of the nanorod itself is identified at $f_1$=1.9 MHz (see inset of Fig. 1b at resonance). At a piezo excitation power of 20 μW the full width at half maximum amounts to FWHM = 300 Hz, leading to a mechanical quality factor Q=6500 ± 2500, defined as Q=f/FWHM. When increasing the excitation power to 126 μW, the resonance develops a hysteresis upon sweeping the drive frequency up and down, revealing a non-linear dynamical behaviour[24].

In order to avoid excess EBD deposition during SEM inspection, which leads to a shift in resonance frequencies, and to allow characterization of the nanorod oscillators in ambient conditions, the above experiments are repeated at room pressure using a diffraction limited high magnification factor (×500) optical microscope. The nanorods scatter light efficiently enough to make them easily visible under the microscope (see Supplementary Fig. A) and the convenience of this technique enables rapid cycling through the characterization of nanorods of successive fabrication batches (see Methods). Even though composition and material properties of EBD-grown carbon-based material are not well controlled, vibrational eigenfrequencies in the MHz range with high Q in air ($10^2$ to $10^3$) are consistently observed on a set of 14 nanorods (see Table 1).

Following the mechanical characterization, the nanorod is positioned into the optical microcavity as depicted schematically in Fig. 2a. The micron-sized Fabry-Pérot cavity consists of two Bragg-coated concave fibre end-facets[25], with a mode waist radius of 3.4 μm (see Methods). The experiment is performed at ambient pressure and room temperature. To manipulate the nanorod in the cavity, the host silicon lever along with the piezo-actuator is glued on a thin copper holder and mounted in turn on a XYZ nano-positioning stack (Attocube ANP100), allowing precise positioning of the nanorod over a range of 5 mm in all three directions. The 2 μm thin silicon lever allows inserting the nanorod into the 42 μm gap between the two micromirrors, thus entering the optical mode region (see Fig. 2b). The optical set-up schematics are shown in Fig. 2c.

The position of the nanorod along the mode modifies the cavity transmission. For each position (x,y) of the nanorod in the plane transverse to the optical axis Z for a given $z_0$, resonant transmission T(x,y) is measured. The 2D plot of T(x,y) is shown in Fig. 3a. Fig. 3b, for comparison, displays a simulation of T(x,y) (see Methods) where the scattering induced by the nanorod is approximated by the convolution of the cavity mode cross-section (Fig. 3c) with the cross-section of the nanorod attached to the end of the microlever (Fig. 3d). The comparison between Figs 3a and 3b allows to determine the optimum (x,y) position of the nanorod in the cavity for subsequent measurements: this position is chosen in order for the nanorod to perturb the cavity mode, while making sure the silicon lever does not contribute much to the perturbation. To this end, we focus on situations where the nanorod is positioned to reduce the resonant transmission by a factor of 2 (see e.g. arrow in Figs 3a and 3b, corresponding to the situation shown in Fig. 3e).

In this case, the standing wave intensity distribution in the resonant cavity can be imaged by moving the nanorod base position $z_0$ along Z while keeping the x- and y-coordinate fixed (see Fig. 2a) and simultaneously recording the cavity transmission and reflection. The resulting graph is shown in Fig 4a. It has the expected λ/2 periodicity. To read out the position fluctuations of the nanorod in the noise spectrum of the transmission with the highest sensitivity, we select a position $z_0$ of maximal gradient $dT/dz_0$ (arrow in Fig. 4a) and lock the cavity on resonance with the laser (see Methods). Data obtained with nanorod 6 (see Table 1) are shown in Figs 4b and 4c. Fig. 4b displays a clear resonance around 13 kHz that is observed 20 dB above the noise floor and which reflects the thermal motion of the host silicon lever. Indeed, the lever vibrational motion translates to a motion of the nanorod base position $z_0$ of small amplitude $<< \lambda/2$, which modulates the cavity transmission in proportion to $dT/dz_0$. Using the value of $dT/dz_0$, the amplitude of the $z_0$ motion can be inferred from the transmission noise spectrum (see Fig. 4b). An excellent fit is obtained using a harmonic oscillator model for the lever flexural thermal motion, which is described by the amplitude component $z_{0,\omega}$ at angular frequency $\omega=2\pi f$ in a frequency window $\delta f$

$$\frac{|z_{0,\omega}|^2}{\delta f} = \frac{k_B T}{K} \frac{\Gamma \omega_0^2}{\left(\omega_0^2 - \omega^2\right)^2 + (\omega\Gamma)^2}, \qquad (1)$$

**Fehler! Textmarke nicht definiert.**

where K is the lever spring constant, $f_0$ its eigenfrequency and $\Gamma$ its damping rate, such that $Q=(2\pi/\sqrt{3})(f_0/\Gamma)$. From the fit, we obtain K=0.21 N/m and $f_0$=13.17 kHz, in agreement with the specifications of the lever, and Q=14, a typical value for an AFM lever at room pressure. At higher frequencies, we observe three resonances: at $f_1$=473 kHz (5 dB above the noise floor), at $f_2$=784 kHz (Fig. 4c inset, 3 dB above the noise floor) and a third at $f_3$=1.172 MHz, which appears as a slight protrusion over the noise floor. They are the three eigenmodes of nanorod 6 already identified by piezo-actuation under the optical microscope (see Table 1). For calibration purposes, the experiment is repeated in the cavity under external piezo-actuation of the nanorod at the same position $z_0$ (arrow in Fig. 4a). We measure the response of the cavity transmission to this externally driven vibrational excitation in the low amplitude limit ($z<< \lambda/2$). Under our experimental conditions, the cavity transmission responds linearly in $dT/dz_0$ and in z. This calibration allows deducing the nanorod first flexural mode vibrational amplitude z (shown in Fig. 4c) from the noise measurement when the external driving is switched off. However, because of the limited resolution of the optical microscope, we estimated this calibration to be valid only within a factor 3. Using the harmonic oscillator model from Eq. (1), we fitted the resonance spectrum of the first thermally driven mode of nanorod 6 and obtained $f_1$=473.4 kHz, $K_{rod}$=1.35 N/m and $Q_1$=215, in perfect agreement with the piezo-actuation experiments.

We infer from Fig. 4c a detection sensitivity of 200 fm/√Hz, limited here by the spectrum analyzer noise level. The shot-noise of light crossing the cavity sets a lower bound to the amplitude of the vibrational fluctuation which can be read-out in the transmission. This bound is $(1/(dP_t/dz)) \times \sqrt{(2h\nu P_t)}$ where $P_t$ is the transmitted optical power and $h\nu$ the energy of an incident photon[26]. In the present experiment, this amounts to 100 fm/√Hz, very close to our observed sensitivity.

This sensitivity can be further improved using a lower noise analyzer, a homodyne detection scheme and most importantly a better cavity. A similar fibre based microcavity of finesse 40 000 has already been demonstrated[25], which corresponds to an improvement of about one order of magnitude in comparison to the cavity used in this work (see Methods). Increased finesse results in an increased value of $dP_t/dx$ (see Supplementary Note A and ref [4]). Additionnally, our present cavity suffers from a slight misaligment limiting its transmission and a 100-fold transmission improvement can likewise be expected in case of perfect alignment[25]. All these improvements should bring the shot-noise limited sensitivity down to the fm/√Hz level for the same nanorod and incident power, which is close to state of the art nanomechanical displacement sensitivity reported using a Single Electron Transistor at mK temperature[16]. Eventually the detection sensitivity will depend on the strength of the light-nanoresonator interaction. For example, when bringing in a carbon nanotube, a resonant dipole interaction of the cavity field with an excitonic line of the tube could be advantageously explored.

Such high sensitivity would allow the vibrational spectroscopy by optical means of nanomechanical resonators of virtually any size or composition. It is also an asset for detecting their mechanical response to weak perturbations in sensing applications, like ultra-sensitive mass sensing[27,28]. Used as a nanosensor, the nano-optomechanical system under investigation could benefit both from high sensitivity to inertial mass accretion, thanks to small mass of the nanorod and its high Q in air ($10^2$ to $10^3$), and from low-noise optical interferometry detection[29-31]. Even more, an improvement of the cavity finesse or a reduction of nanomechanical resonator mass would also place the system in a regime of strong optomechanical back-action[32]. With a single wall carbon nanotube positioned in the cavity presented here, we should observe optical self-cooling of the nanotube as well as its optically pumped self-oscillation[4,33]. Such optomechanical control combined with high optical sensitivity will eventually allow operating such nano-optomechanical sensors approaching the limit of Heisenberg fluctuations.

*Methods:*

EBD nanorod grown on a AFM cantilever

The nanorods investigated here are commercially available for AFM applications from Nanotools, Munich, Germany. *www.nanotools.com.* The host AFM lever is from Nanosensors. PPT-CONTR, nominal values: thickness=2 μm, mean width= 50 μm, length= 450 μm, force constant= 0.2 N/m, resonance frequency= 13 kHz. Table 1 contains data acquired by ultrasonic piezo-actuation of the nanorods flexural modes in the x and z directions, imaged under an optical microscope. These data are taken under air atmosphere on 5 distinct fabrication batches. Frequencies $f_1$, $f_2$ and $f_3$ are the three lowest vibrational frequencies identified when performing the experiment. They correspond to flexural modes either in the x-direction (noted -x) or in the z-direction (-z). As exemple, at 66 mW excitation power, the extremity of nanorod 6 (batch 3) oscillated in air with an amplitude z of about 1 μm, 0.5 μm and 0.25 μm for the three identified lowest frequency resonances, respectively. On nanorod 12, we performed piezo-actuation both of the x- and z-modes (see Supplementary Fig. A for pictures of the x-mode). Because the nanorod's shape is almost cylindrical, the x and z modes are in the same frequency range. They also have similar Q. The significant variation of the first flexural mode frequency from one nanorod to another is not due to geometrical variation. Indeed, the rods' dimensions were systematically measured in the SEM and showed a distribution of less than 15% in length and 20% in diameter. Moreover, using elasticity theory and assuming an homogeneous material, we could not model the three resonances observed on nanorod 6, be it with numerical means or with an analytical formula for a cone[34]. Possibly the EBD growth process is responsible for material inhomogeneity, depending on exact fabrication conditions.

|  | Nanorod | $f_1$(MHz) | $Q_1$ | $f_2$(MHz) | $Q_2$ | $f_3$(MHz) | $Q_3$ |
|---|---|---|---|---|---|---|---|
| Batch 1 | nanorod 1 | 1.90 (-x) | 6500±2500 (vacuum) | | | | |
|  | nanorod 2 | 1.67 (-x) | 170 ±100 | | | | |
|  | nanorod 3 | 1.88 (-x) | 1900±500 | | | | |
| Batch 2 | nanorod 4 | 0.936 (-x) | 460 ± 250 | 1.67 (-x) | | | |
| Batch 3 | nanorod 5 | 0.363 (-x) | 190 ± 90 | 0.601 (-x) | 260 ± 100 | | |
|  | nanorod 6 | 0.474 (-z) | 250 ± 100 | 0.784 (-z) | 250 ± 150 | 1.172 (-z) | 250 ± 100 |
| Batch 4 | nanorod 7 | 1.706 (-z) | 630 ± 300 | 4.375 (-z) | 1200 ± 450 | | |
|  | nanorod 8 | 1.072 (-z) | 390 ± 150 | 1.497 (-z) | 340 ± 200 | | |
|  | nanorod 9 | 1.162 (-z) | 360 ± 200 | | | | |
|  | nanorod 10 | 1.138 (-z) | 570 ± 250 | | | | |
|  | nanorod 11 | 1.300 (-z) | 500 ± 200 | | | | |
| Batch 5 | nanorod 12 | 1.059 (-z) | 415 ± 100 | 1.479 (-z) | 525 ± 150 | 1.677 (-x) | 930 ± 300 |
|  | nanorod 13 | 0.981 (-z) | 630 ± 250 | 1.371 (-z) | 480 ± 250 | | |
|  | nanorod 14 | 1.691 (-z) | 420 ± 200 | | | | |

**Table 1. Mechanical resonances of the nanorods characterized at room pressure by ultrasonic actuation using optical microscope imaging.** Nanorod 1 was investigated under vacuum in the SEM. (-x) and (-z) indicate the direction of the nanorod vibration.

Fibre-integrated cavity set-up

Our optical resonator is a micron-sized fibre-integrated Fabry-Pérot cavity designed for a wavelength of 780 nm. The cavity mirrors are two optical fibre end-facets (input and output of the cavity) formed by $CO_2$ laser machining in order to obtain a concave transverse profile and then coated with dielectric multiple layers to provide high reflectivity. The two resulting high quality mirrors face each other at a distance of 42 μm (Fig. 2b), leading to a stable cavity mode with a waist radius of 3.4 μm. The mirrors' distance can be tuned using shear piezo-plates holding the two fibres. The incoupling fibre is a single-mode fibre to get a defined mode profile for good mode matching and the outcoupling fibre a gradient index fibre. An external cavity laser diode[35] is run monomode at 780 nm with a line width in the MHz range and coupled into the input fibre (Fig. 2c). The laser is frequency-stabilized on a Rubidium atomic resonance. Two Faraday isolators (Linos, 30 dB and 40 dB) mounted in series provide isolation of the laser. The cavity transmission is collected after the output fibre on a pre-amplified photodiode (Thorlabs PDA55). A half wave-plate placed before the input fibre is used to select one of the two linearly polarized ground modes of the cavity. We swept the cavity length at 3 Hz over 2 free-spectral ranges (FSR) while monitoring the transmission and compared the FSR with the cavity resonance width, leading to a measured finesse of F=5000. A finesse of F=40 000 can currently be reached with this cavity technology[25]. A lower value of F is observed if the cavity is slightly misaligned or contaminated. Cavity transmission is also reduced in case of misalignment because of reduced mode matching. We designed the whole set-up arrangement in a compact way in order to place it in a centimetre-sized vacuum glass cell and operation under ultra-high vacuum will be possible at a later time.

Simulation of T(x,y):

As discussed in ref [4], the scattering by the needle-shaped nanorod in the cavity mode is well approximated by that of an homogeneous plane with a scattering parameter $\Sigma=\Sigma_1+i\Sigma_2$. $\Sigma$ increases linearly with the number of point dipoles composing the needle which are positioned in the optical mode. Hence $\Sigma$ is a function of the nanorod position (x,y) and is proportional to the overlap integral between the cross-section of the scatterer (Fig. 3d) and the cavity mode profile (Fig. 3c). In ref [4], an expression for the cavity transmission as a function of $z_0$, cavity finesse F ($g=2F/\pi$) and $\Sigma_1$ is derived in the limit of $\Sigma_1 \ll g$ and z=0:

$$T(z_0) = \frac{1}{\left(1+2g\Sigma_1 \sin^2(kz_0)\right)^2} \quad (2)$$

Keeping $z_0 \neq 0$ and g constant, the transmission T(x,y) is simulated using equation (2) and taking $\Sigma_1(x,y)$ to be proportional to the overlap integral. The result of the simulation is shown in Fig. 3b. The comparison between Figs 3a and 3b allows estimating the nanorod position (x,y) in the cavity mode. The resolution of this procedure is limited by the transverse size of the cavity mode but the non-linearity in $\Sigma_1$ of equation (2) helps improving this resolution.

Cavity stabilization:

The fibre ends forming the mirrors of the cavity are glued on shear piezo elements. By applying the oscillator output voltage of a lock-in amplifier (Signal Recovery 7265) at 97 kHz to one of these shear piezos, the cavity length is modulated around a cavity resonance. 10 percent of the photodiode signal of the light transmitted through the cavity is sent back on the lock-in, mixed with its reference and low-pass filtered to produce a dispersive error signal. It is null at cavity resonance and its sign has a one to one correspondence with the sign of the cavity detuning. This error signal is fed into an integral-proportional lock-box. The lock-box output voltage is low-pass filtered, amplified and applied to the shear-piezo holding the second cavity mirror, to provide feedback on the cavity length.

Acknowledgments:

We gratefully acknowledge financial support of the Alexander von Humboldt Foundation, the German-Israeli Foundation (G.I.F.), the German Excellence Initiative via the Nanosystems Initiative Munich (NIM) and the Center for NanoScience.

The authors declare no competing financial interest.

**Figure 1 Ultrasonic actuation of the nanorod flexural resonance.** a) Micrograph of the AFM cantilever with a nanorod at its extremity. b) Main panel: SEM picture of an EBD nanorod grown at the end of an AFM lever. Inset: Piezo-actuated vibrational resonance of nanorod 1 flexural ground mode at 1.9 MHz imaged in the SEM.

**Figure 2 Nanorod vibrating in the microcavity resonantly probed by a laser.** a) Schematics of nanorod at position $z_0$ in the microcavity and vibrating with an amplitude z. b) Optical micrograph of the host silicon lever plunged between the two fibre end-facets in order to position the nanorod in the cavity mode. c) Set-up schematics (PBS: polarizing beam splitter, FC: fibre coupler, PD: photodiode).

**Figure 3 In situ positioning of the nanorod in the cavity mode.** a) Imaging of the nanorod in the cavity mode trough the cavity transmission T(x,y). The arrow indicates the operating point where the resonant transmission is reduced by a factor 2. b) Simulated cavity transmission T(x,y) (see Methods) (the pixels show the grid taken for computation). Arrow at the same position as in a). c) Cross-section of a Gaussian mode intensity distribution representing the cavity mode. d) Planar section of the nanorod placed at the end of the AFM lever. e) Nanorod position in the cavity mode, corresponding to the arrow in a) and b).

**Figure 4 Perturbation of the cavity transmission by the nanorod.** a) Optical power transmitted and reflected by the cavity at resonance as a function of the nanorod base position $z_0$ normalized to the laser wavelength $\lambda$ ($z_s$ being the average position of the lever, roughly in the middle of the cavity). b) Brownian motion amplitude spectrum of the AFM lever extremity $z_0$ holding the nanorod, from noise measurement taken at maximum gradient $dT/dz_0$. c) Brownian motion amplitude spectrum z for the first flexural resonance of nanorod 6. Inset: Transmission noise power spectrum around the frequency of the second flexural resonance of nanorod 6, for a resolution bandwidth of 300 Hz of the spectrum analyser.

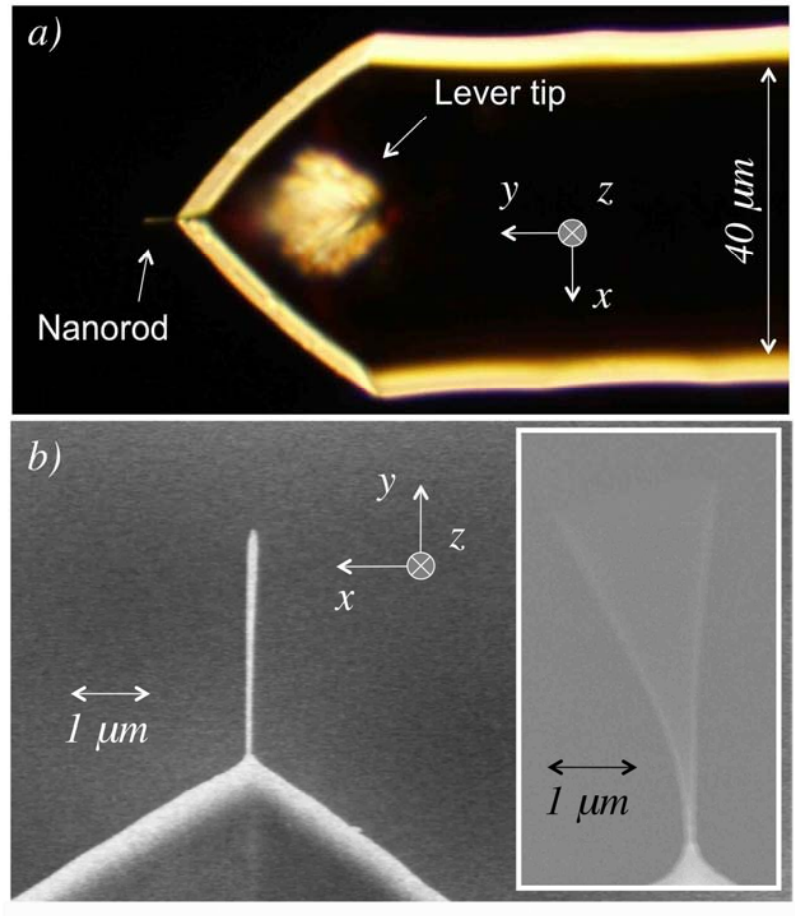

Figure 1

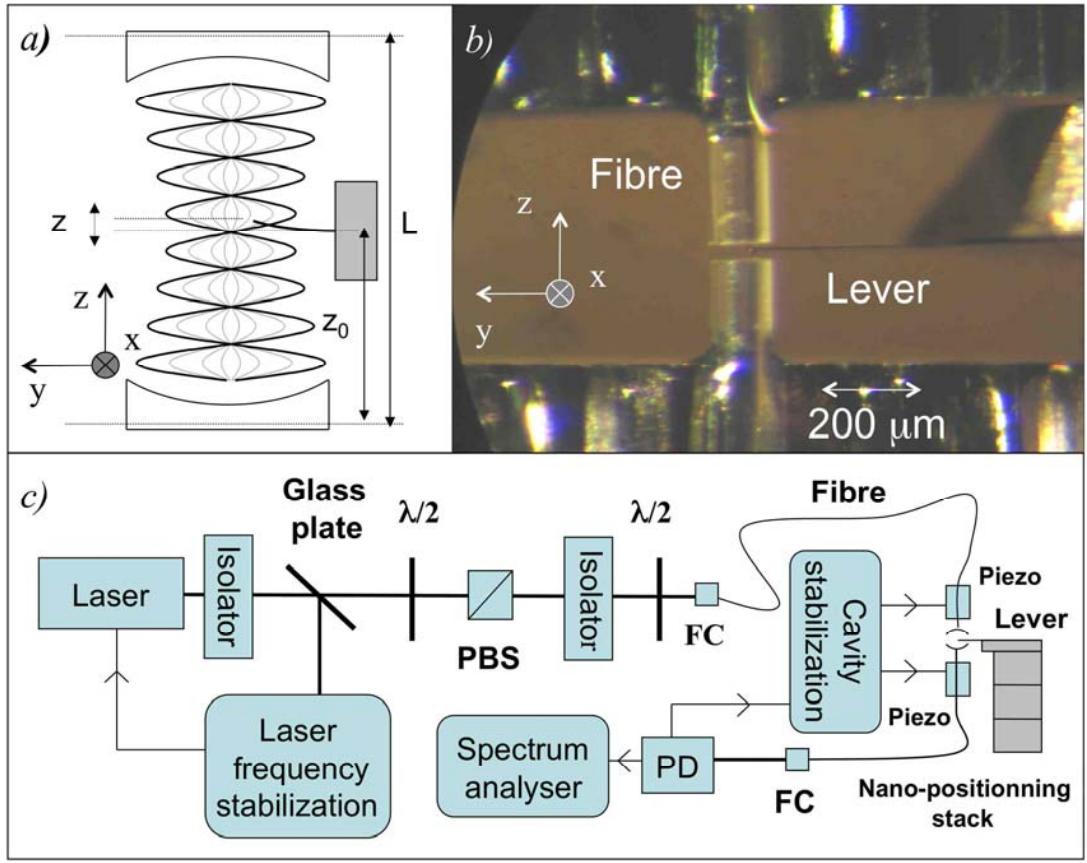

Figure 2

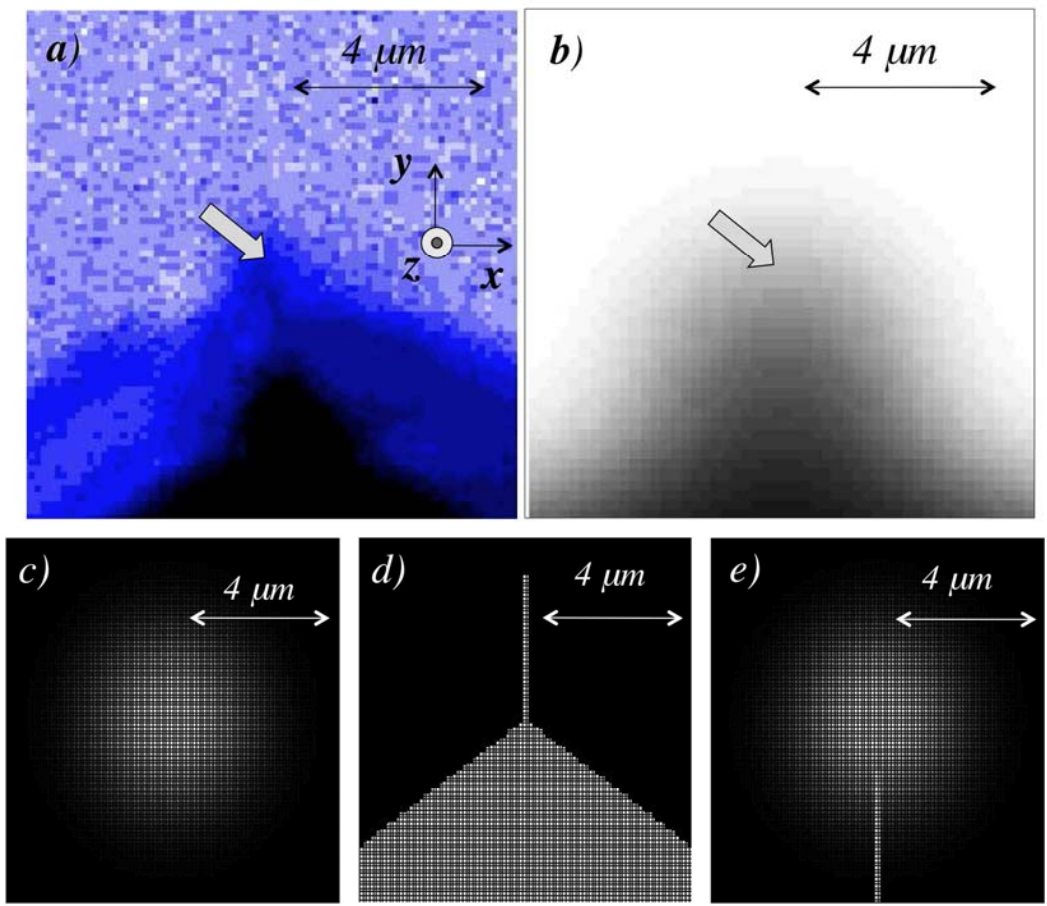

Figure 3

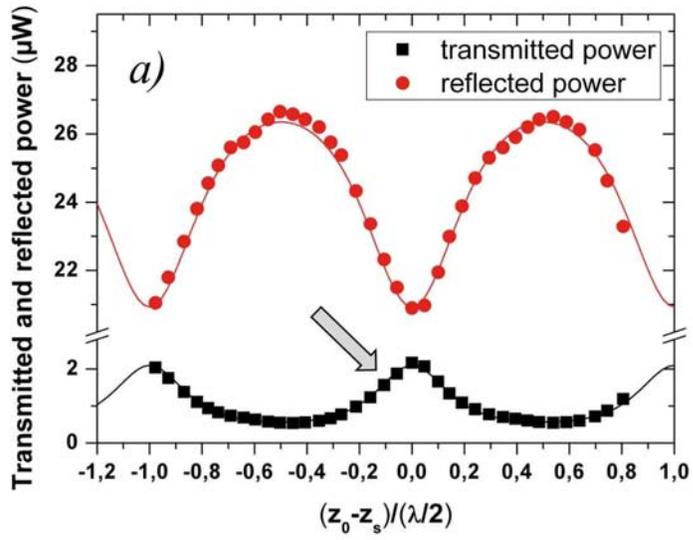
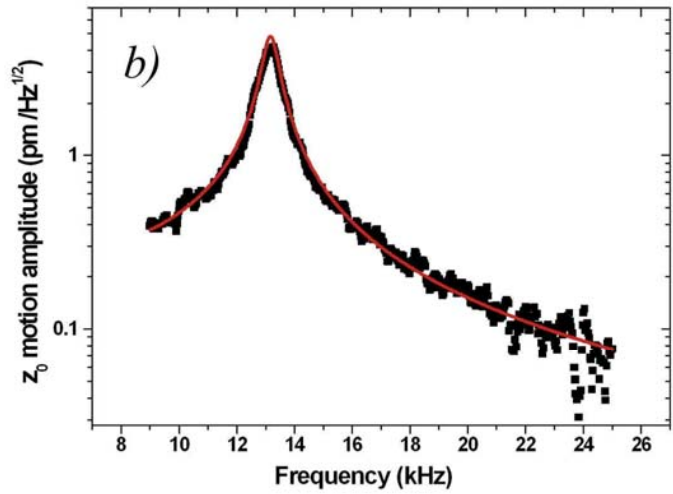
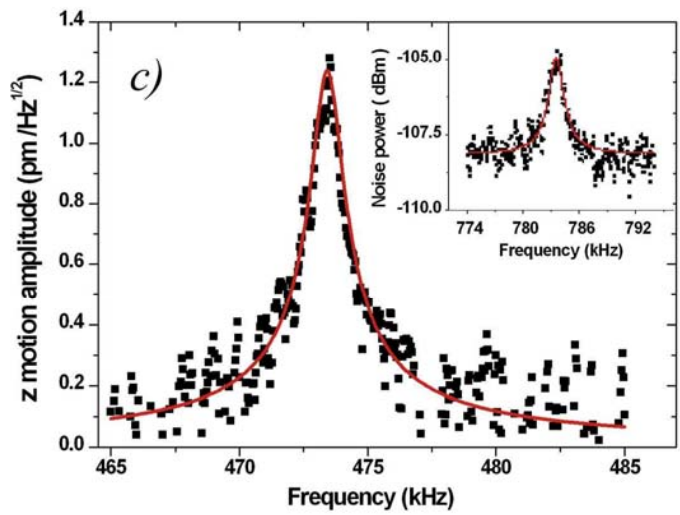

Figure 4